\documentclass{article} 

\title{The solutions to single-variable polynomials, implemented and verified in \Lean\\
{\normalsize An experience report on learning how to use a computer proof assistant}
}

\author{%
Nicholas Dyson\thanks{%
University of Birmingham, United Kingdom}
\and
Benedikt Ahrens\thanks{%
Delft University of Technology, The Netherlands, and University of Birmingham, United Kingdom,
email:~\email{B.P.Ahrens@tudelft.nl}}
\and
Jacopo Emmenegger\thanks{
University of Genoa, Italy,
email:~\email{emmenegger@dima.unige.it}}
}
\date{}

\usepackage{amsmath,amsthm}
\usepackage{graphicx}
\usepackage[combine]{ucs}
\usepackage{xcolor}

\definecolor{keywordcolor}{rgb}{0.7, 0.1, 0.1}   
\definecolor{commentcolor}{rgb}{0.4, 0.4, 0.4}   
\definecolor{symbolcolor}{rgb}{0.0, 0.1, 0.6}    
\definecolor{sortcolor}{rgb}{0.1, 0.5, 0.1}      
\definecolor{errorcolor}{rgb}{1, 0, 0}           
\definecolor{stringcolor}{rgb}{0.5, 0.3, 0.2}    

\usepackage{listings}

\lstdefinestyle{leannocolor}{
   basicstyle={\ttfamily},
   identifierstyle={\ttfamily},
   keywordstyle=[1]{\ttfamily},
   keywordstyle=[2]{\ttfamily},
   keywordstyle=[3]{\ttfamily},
   stringstyle={\ttfamily},
   commentstyle={\ttfamily}
}

\usepackage{xargs}
\usepackage{xifthen}

\usepackage{xspace}

\usepackage{bookmark}
\usepackage{hyperref}
\hypersetup{
    colorlinks=true
   }
\usepackage[capitalise]{cleveref}

\usepackage[draft]{optional}
\usepackage[colorinlistoftodos,prependcaption,textsize=tiny]{todonotes}
\newcommandx{\plan}[1]{}
\newcommandx{\BA}[1]{}
\newcommandx{\ND}[1]{}
\newcommandx{\JE}[1]{}
\opt{draft}{
   \renewcommandx{\plan}[1]{\todo[color=blue!30]{Plan: #1}\PackageWarning{TODO}{Plan: #1}}
   \renewcommandx{\BA}[1]{\todo[color=orange!30]{BA: #1} \PackageWarning{TODO}{BA: #1}}
   \renewcommandx{\ND}[1]{\todo[color=green!30]{ND: #1}\PackageWarning{TODO}{ND: #1}}
   \renewcommandx{\JE}[1]{\todo[color=red!30]{JE: #1}\PackageWarning{TODO}{JE: #1} }
}

\newcommand{\linkprefix}{id:}
\newcommand{\coqdocbaseurl}{https://NicholasDyson.github.io/non-mathlib/}
\newcommand{\urlhash}{\#}
\newcommand{\coqdocurl}[2]{\coqdocbaseurl #1.html\urlhash #2}
\newcommand{\nolinkcoqident}[1]{\nolinkurl{#1}} 
\makeatletter
\newcommand{\coqident}{\begingroup\@makeother\#\@coqident}
\newcommand{\@coqident}[3][]{%
  \ifthenelse{\isempty{#2}}%
  {\nolinkcoqident{#3}}%
  {\ifthenelse{\isempty{#1}}%
  {\href{\coqdocurl{#2}{#3}}{\color{blue}{\nolinkcoqident{\linkprefix #3}}}}%
  {\href{\coqdocurl{#2}{#3}}{\nolinkcoqident{#1}}}}%
\endgroup}

\theoremstyle{plain}
\newtheorem{thm}{Theorem}[section]
\newtheorem{prop}[thm]{Proposition}

\theoremstyle{definition}

\newtheorem{rmk}[thm]{Remark}

\newcommand{\Lean}{\textsf{Lean}\xspace}
\newcommand{\mathlib}{\textsf{mathlib}\xspace}
\newcommand{\email}[1]{\href{mailto:#1}{\texttt{#1}}}

\begin{document}

\maketitle

\begin{abstract}
 In this work, we describe our experience in learning the use of a computer proof assistant---specifically, \Lean---from scratch,
 through proving formulae for the solutions of polynomial equations.
 
 Specifically, in this work we characterize the solutions of quadratic, cubic, and quartic polynomials over certain fields, specifically,
 fields with operations returning square and cubic roots of characteristic other than two or three.
 
 The purpose of this work is thus twofold.
 Firstly, it describes the learning experience of a starting \Lean user, including a detailed comparison between our work in \Lean and very closely related work in Coq.
 Secondly, our results represent a modest improvement over the aforementioned related work in Coq, which we hope will be of some scientific interest.
\end{abstract}


\section{Introduction}
\label{sec:intro}

This work arose from the final year project of one of the authors (N.~Dyson).
They expressed an interest in learning computer theorem proving, and, after studying Freek Wiedijk's list of 100 theorems~\cite{100_theorems}, decided to work towards a formal certification of the solutions of cubic and quartic polynomials, and, possibly, a proof of the non-existence of a closed formula for quintic polynomials.

The latter goal turned out to be too ambitious to fit into the project period. 
However, the other goals were achieved, and in the process, we learned some things we would like to share in this article.

Specifically, we present in this work a formal proof, in the computer proof assistant \Lean, of the correctness of closed formulas for the solutions of cubic and quartic polynomials.

In contrast to existing work (see also~\cref{sec:relwork}), our solution is implemented for an arbitrary field, parametrized by functions for computing square roots and cube roots.

In the remainder of this introduction, we present the mathematical results that we have formalized in \Lean. 
In~\cref{sec:relwork} we compare our approach to previous work on the topic of cubic and quartic polynomials.
The structure of the rest of this article is explained in~\cref{sec:synopsis}.

\subsection{Polynomial equations}
\label{sec:intro:poly}

Polynomial equations (in a single variable $x$) are equations where a sum of integer powers of $x$, each multiplied by a coefficient, add to make zero.
The most well-known examples are the quadratic equations:
$ax^2 + bx + c = 0$, where $a$, $b$, and $c$ are the coefficients and the powers of $x$ are $2$, $1$, and $0$.
These logically progress to the cubics, with an $x^3$ term, and the quartics, with an $x^4$ term, and beyond.
The solution to a general quadratic has been known since antiquity, but the solutions to general cubic and quartic equations are somewhat more recent,
being devised by the Italian mathematicians Cardano and Ferrari during the sixteenth century.

Formalised proofs of these results already exist in various computer proof assistants, but not in \Lean (beyond the quadratic).
In order to have an all-round experience,
we have implemented the solutions to cubic and quartic equations
in two ways---once from scratch and without using any preexisting libraries, and again using \mathlib, which is the de-facto standard library of \Lean.
We did this without reference to preexisting formalisations in different proof assistants,
but a comparison between our version and the existing Coq version is given in~\cref{sec:coqcomp}.

\subsection{Solutions to polynomial equations}
\label{sec:intro:mathback}

Here we provide derivations of the formulae for the solutions
to quadratic, cubic and quartic equations based on
the presentation in~\cite{cardano, brookfield_factoring_quartic_polys}.

We work in a field $k$ of characteristic different from 2 and 3.
We assume that $k$ has a choice of a
square root and a cube root for each of its elements.
This holds for any algebraically closed field,
and for the field of complex numbers in particular.
We write $\sqrt{x}$ and $\sqrt[3]{x}$ for the chosen roots of $x$.
Note also that, given a cube root $s$ of $x$,
the other two cube roots of $x$ are
\[
\frac{-1 + \sqrt{-3}}{2}\, s
\qquad \text{and} \qquad
\frac{-1 - \sqrt{-3}}{2}\, s.
\]
The leading coefficient of a polynomial,
\textit{i.e.}\ the coefficient of the highest power,
is non-zero by definition.
Thus we assume without loss of generality
that the leading coefficient is 1.

\subsubsection{Quadratic}
\label{sec:intro:quadratic}

Although the formula for the solutions
to a quadratic equation is well-known,
its derivation illustrates the general technique
of substituting variables to reduce to simpler or known cases.

We can rewrite $x^2 + bx + c = 0$ as
\[
\left(x + \frac{b}{2}\right)^2 - \frac{b^2}{4} + c = 0
\]
which is an equation of the form $y^2 - d = 0$.
The solutions to the latter are clearly $\pm \sqrt{d}$,
thus
\begin{equation}\label{quadr-sol}
x = \frac {-b \pm \sqrt{b^2 - 4c}}{2}.
\end{equation}

\subsubsection{Cubic}
\label{sec:intro:cubic}

The formula for the solutions to a generic cubic equation is less well-known.
First, observe that we can reduce a general cubic to an equivalent cubic with no $x^2$ term, usually referred to as a \emph{depressed cubic}, using a change of variable.
Given $x^3 + bx^2 + cx + d = 0$, let $x = u - \frac{b}{3}$.
Then we have
\[
u^3 + \left(-\frac{b^2}{3} + c\right)u
	+ \left(\frac{2b^3}{27} - \frac{bc}{3} + d\right)
=0.
\]
If we can solve this depressed cubic,
we can clearly recover the solutions to the original cubic
by $x = u - \frac{b}{3}$.

We describe here Cardano's original method~\cite{cardano}%
\footnote{It is generally recognised, although not entirely uncontroversial~\cite{guilbeau_cardano}
  that Cardano was the first person to discover this method.}
to solve a depressed cubic $ u^3 + cu + d = 0 $.
Assume that $d\neq 0$ and let $s,t$ be such that
\begin{equation}\label{cubic_sims}
\begin{cases}
3st = c
\\
t^3 - s^3 = d.
\end{cases}
\end{equation}
Then at least one among $s$ and $t$ is not zero,
and $s - t$ is a solution to the original equation
as it can be easily verified.
Let us suppose that $s \neq 0$.
Thus $t = \frac{c}{3s}$ and
from the second equation in~\eqref{cubic_sims} we obtain
\[
\left(s^3\right)^2 + ds^3 - \frac{c^3}{27} = 0
\]
which we can solve for $s^3$ using~\eqref{quadr-sol}.
It follows that
\begin{gather}
\label{eq:cubic-system-sol1}
s = \sqrt[3]{-\frac{d}{2} + \sqrt{\frac{d^2}{4} + \frac{c^3}{27}}}
\\[1ex]
\label{eq:cubic-system-sol2}
t = \sqrt[3]{\frac{d}{2} + \sqrt{\frac{d^2}{4} + \frac{c^3}{27}}}.
\end{gather}
and Cardano's formula for a solution
to the depressed cubic equation $u^3 + cu + d = 0$ is
traditionally presented as
\begin{equation}\label{eq:depr-cubic-sol}
u = \sqrt[3]{-\frac{d}{2} + \sqrt{\frac{d^2}{4} + \frac{c^3}{27}}}
- \sqrt[3]{\frac{d}{2} + \sqrt{\frac{d^2}{4} + \frac{c^3}{27}}}.
\end{equation}

A naive application of this formula is not guaranteed to work though.
Indeed, we can compute the values of $s$ and $t$
since our field $k$ has a choice of a cube root for each element.
However this choice need not respect the condition $3st=c$
from~\eqref{cubic_sims}.

One approach, that we follow, is to use only one among $s$ and $t$
and to compute the other one using the condition $3st=c$.
In this approach some care has to be taken to avoid division by zero.
As we assume $d \neq 0$, note that $s$ and $t$ cannot be both zero,
and one of them is zero if and only if $c=0$.
In this case the solutions are just the cube roots of $-d$.
Otherwise, we have the following.

\begin{prop}\label{prop:depr-cubic-sol}
Let $k$ be a field of characteristic different from 2 and 3,
together with a choice of a square root and a cube root for each element.
Consider the depressed cubic equation $u^3 + cu + d = 0$,
where $c,d\neq0$, and let $s$ be as in~\eqref{eq:cubic-system-sol1}.
Then a solution to the depressed cubic equation is
\[
u = s - \frac{c}{3s}.
\]
The other solutions can be found by replacing $s$
with the other two cube roots of
$-\frac{d}{2} + \sqrt{\frac{d^2}{4} + \frac{c^3}{27}}$.
\end{prop}
In case we start with a generic cubic equation
$ax^3 + bx^2 + cx + d = 0$,
the hypothesis of \cref{prop:depr-cubic-sol} must hold for the
associated depressed cubic and translate as
\begin{equation}\label{eq:cubic-hyp}
3ac - b^2 \neq 0
\qquad\text{and}\qquad
2b^3 - 9abc + 27a^2d \neq 0.
\end{equation}
A formula for the solutions to a generic cubic equation
satisfying~\eqref{eq:cubic-hyp} can then be derived from
\cref{prop:depr-cubic-sol},
although it is not shown here for reasons of space.

\begin{rmk}
\Cref{prop:depr-cubic-sol} only allows us to find solutions
to certain cubic equations, namely those satisfying conditions~\eqref{eq:cubic-hyp}.
A global formula that covers also the other cases
would be easy to implement by case distinction using classical logic.
In order to implement such a global formula while staying constructive
one would probably need to consider a suitable formulation of field,
as a field with an apartness relation~\cite{10.1305/ndjfl/1093635926}.
On the other hand, in case one (possibly both) of the two expressions
in~\eqref{eq:cubic-hyp} is zero, the solutions are easy to find.

The need for excluding a certain number being zero appears to be unavoidable; all formalizations known to us either require that it is decidable whether a value is equal to zero or not,
or simply assert that the quantity is not equal to zero (see \cite{coq_proof} and \cite{metamath_proof} for two examples).

\end{rmk}


\subsubsection{Quartic}
\label{sec:intro:quartic}

Once the solutions to the general cubic are known,
these can be used to find the solution to the general quartic
via an elegant derivation~\cite{brookfield_factoring_quartic_polys}.
We can perform a similar reduction to the way we reduced the cubic to a depressed cubic, this time eliminating the $x^3$ term.

Let $x = u - \frac{b}{4}$, substituting this value
in the general quartic $x^4 + bx^3 + cx^2 + dx + e = 0$
we obtain
\begin{equation}
u^4 + \left(c -\frac{3b^2}{8}\right)u^2 + \left(\frac{b^3}{8} - \frac{bc}{2} + d\right)u  + \frac{b^2c}{16} - \frac{3b^4}{256} - \frac{bd}{4} + e
\end{equation}
%
If we can solve this depressed quartic, we can recover the solution to the general quartic by subtracting $\frac{b}{4}$.

To see a route to a solution,
recall that, by the Fundamental Theorem of Algebra,
this depressed quartic (and indeed, any quartic equation) must be the product of four linear expressions.
It follows that every quartic can be expressed
as a product of two quadratic polynomials:
\begin{equation}\label{quartic-decomp}
u^4 + cu^2 + du + e = (u^2 + pu + q)(u^2 + ru + s)
\end{equation}
for some $p$, $q$, $r$, and $s$.
If we can find a method to determine suitable values for $p$, $q$, $r$, and $s$, we will have solved the equation, as the solution to a quadratic is already known.
Note that the decomposition into a pair of quadratics will not be unique, as there are three different ways to pair the four linear expressions with each other.
However, the solutions are clearly independent of the chosen decomposition.

Unfolding the right-hand product in~\eqref{quartic-decomp}
we obtain the system of equations
\begin{equation}\label{quartic-system}
\begin{cases}
p + r = 0
\\
q + s + pr = c
\\
ps + qr = d
\\
qs = e
\end{cases}
\end{equation}
%
If we find a solution to this set of equations,
we have determined the quadratic decomposition of our depressed quartic.

Note first that, if $p=0$, then $r=0$
and the original depressed quartic is of the form
\[
\left(u^2\right)^2 + c u^2 +e
\]
which can be solved using
the formula for the solutions to a quadratic~\eqref{quadr-sol}.
In particular, 
\[
p = 0 \quad \text{if and only if} \quad d=0.
\]
Assuming $p \neq 0$,
the system~\eqref{quartic-system}  reduces to the equation
\begin{equation}\label{resolvent_cubic}
\left(p^2\right)^3 + 2c\left(p^2\right)^2 +\left(c^2 - 4e\right)p^2 - d^2 = 0
\end{equation}
which is sometimes referred to as
the \emph{resolvent (cubic)}~\cite{brookfield_factoring_quartic_polys} of the quartic.
As we already know how to solve cubic equations,
we can find a value for $p^2$ and, in turn, solve the system~\eqref{quartic-system}.
Of course, there are three possible choices for $p^2$,
and these correspond to the three different ways to
factor a depressed quartic into a product of quadratics.
Thus we have proved the following.

\begin{prop}\label{prop:depr-quart-sol}
Let $k$ be a field of characteristic different from 2 and 3,
together with a choice of a square root and a cube root for each element.
Consider the depressed quartic equation $u^4 + cu^2 + du + e = 0$,
where $d,e\neq0$.
Then the four solutions to the depressed quartic equation
can be computed as the solutions to the quadratic equations
\begin{gather*}
u^2+pu+ \frac {p^2 + c + \frac{d}{p}}{2} = 0
\\
u^2 -pu + \frac {p^2 + c - \frac{d}{p}}{2} = 0
\end{gather*}
where $p^2 \neq 0$ is a root of the resolvent cubic~\eqref{resolvent_cubic}.
\end{prop}

In the implementation,
a solution to the resolvent cubic~\eqref{resolvent_cubic}
is computed using \cref{prop:depr-cubic-sol}.
So in addition to $d,e \neq 0$ we also assume
\begin{equation}\label{eq:quart-resolv-cond}
c^2 + 12e \neq 0
\end{equation}
as a condition to the formula for the roots of a quartic.

Starting from a generic quartic $x^4 + bx^3 + cx^2 + dx + e = 0$,
the hypothesis of \cref{prop:depr-quart-sol} must hold for the
associated depressed quartic and translate as
\[
\frac{b^3}{8} - \frac{bc}{2} + d \neq 0
\quad \text{and}\quad
\frac{b^2c}{16} - \frac{3b^4}{256} - \frac{bd}{4} + e \neq 0
\]
whereas condition~\eqref{eq:quart-resolv-cond}
translates as
\[
c^2 - 3bd + 12e \neq 0.
\]

Every step of the preceding algorithm is well-defined, such that
the whole procedure could be made into a single formula,
but it would require too much space to be displayed here.
More practically, such a formula would be also very difficult
to fully comprehend.
The product-of-quadratics reduction, by contrast,
is reasonably intuitive and it is possible to logically break down
into smaller formulae that can be individually verified.

\section{Related work}
\label{sec:relwork}


Freek Wiedijk~\cite{100_theorems} maintains a list of one hundred mathematical results and pointers to computer-checked statements and proofs of these results, in various computer proof assistants.
The solutions of a cubic (Problem \#37) and of a quartic (Problem \#46) are among them.
Below, we comment on the solutions of a quartic pointed to in Wiedijk's list; we are not aware of any solutions that are not mentioned in the list, besides our own.

Formulas for the solutions of cubic and quartic equations
have already been formalised in Coq~\cite{coq_proof} which,
among the existing proof assistants,
is the closest one to \Lean,
being also based on dependent type theory.
However, we did not refer to this existing work in our implementation in \Lean,
in order to avoid duplicating the same argument
and to better explore the strengths of \Lean.
We compare our formalisation with the one existing in Coq
in \cref{sec:coqcomp}.

Besides the solution in Coq, \cite{100_theorems} lists solutions in
HOL Light, by John Harrison~\cite{quartic-hol-light},
Isabelle, by Amine Chaieb~\cite{quartic-isabelle},
Metamath, by Mario Carneiro~\cite{quartic-metamath},
and
Mizar, by Marco Riccardi~\cite{quartic-mizar}.
Common to all these solutions, including the one in Coq, is that they apply specifically to polynomials in the complex numbers, not to polynomials over a general field.

Formulas for the solutions of cubic and quartic equations
appear not to have been formalised in \Lean according to~\cite{100_theorems},
and this partly motivated the choice of \Lean over other proof assistants.
Lean is also fairly modern,
first appearing in 2015~\cite{lean_description}
compared to systems like Coq and Mizar which predate even the 1990s, and it supports multiple approaches to proofs and various kinds of automation.

The main \Lean library \mathlib already contains a proof
of the solution to a quadratic equation~\cite{mathlib_quadratic},
proved for a general field with the additional assumption that certain square roots exist (similar to our hypothesis that square and cube roots exist.)
We did not refer to this proof when implementing our own version,
as the point of doing the quadratic in the first place was to 
acquire familiarity with \Lean.
We did refer to it in \cref{sec:stdlib}
when making a second version of the cubic and quartic proofs
that did use \mathlib.

%

\section{Synopsis}
\label{sec:synopsis}

The basic steps of our implementation in \Lean of solutions to
quadratic, cubic and quartic equations are as follows:
\begin{enumerate}
\item\label{step:sub}
Define substituting a value into a general quadratic equation.
\item\label{step:sol}
Define functions that return the two solutions to a quadratic,
given the coefficients of the quadratic as the input arguments.
\item
Prove that the values returned by these functions always make
the quadratic equal to zero when substituted in.
\item\label{step:uniq}
Prove that these solutions are unique;
more precisely, prove that if $x$ is different from the solutions obtained in step~\ref{step:sol}, then $x$ is not a solution.
\item
Repeat steps~\ref{step:sub}--\ref{step:uniq}
for the three solutions of a cubic.
\item
Repeat steps~\ref{step:sub}--\ref{step:uniq}
for the four solutions of a quartic.
\end{enumerate}

Although conceptually simple,
these steps rely on a specific implementation of the arguments
to an equation, \textit{i.e.}\ of algebraic structures of numbers.
In \cref{sec:firstproof} we present a first attempt 
based on a direct implementation of integers and rationals.
This attempt turned out to be unsuccessful and
we discuss the difficulties that arise.

A second approach consists in using an arbitrary field
which is assumed to have square and cube roots,
as these are needed to compute our formulas for the depressed cubic.
Note that this assumption is necessary
if we want our formulas to always compute a solution---%
that is, the assumption is always satisfied by fields
where every square, cubic and quartic polynomial has at least a root.
In \cref{sec:version_two} we present a successful approach
to the problem of finding formulas for the solutions of cubic and quartic polynomial equations
based on an implementation of such a structure as a type class.
Besides the standard components of the field structure,
it includes two functions
that assign to each element $x$ elements $\sqrt{x}$ and $\sqrt[3]{x}$
such that $\sqrt{x}^2 = x = \sqrt[3]{x}^3$.
As a by-product, our formulas for the solutions to cubic and quartic equations are in fact parametrised
over the functions that compute square and cube roots.

Based on this approach, we are able to perform all the above steps.
However, some subtleties arise in \cref{sec:ver_two:cubic}
because of the specific implementation of field that we use.
\Cref{sec:stdlib} discusses a third approach that is based on
Lean's standard library~\cite{lean_mathlib}
and compares it with the one from \cref{sec:version_two}.
In \cref{sec:coqcomp} we compare our implementation, specifically
the version described in \cref{sec:stdlib} based on \mathlib,
with an already existing implementation in Coq of formulas
for the roots of cubic and quartic polynomials~\cite{coq_proof}.
\Cref{sec:concl} contains some concluding remarks.

\paragraph*{Note on the computer-checked code} 
Throughout the paper, \Lean code is printed in distinguished font, e.g., \lstinline!theorem foo : bar!. 
Furthermore, we frequently insert links to an HTML version of our library, hosted on Github, as in \coqident{field_definition}{square}. 
In such links, we add the prefix ``\linkprefix'' before the \Lean identifier.
The HTML version itself presents an interface to the Lean code, in the style of the \mathlib library. From that interface, there are links to the actual source code of our library, for instance, to the full definition of the \href{https://github.com/anonymousLeanDocsHosting/lean-polynomials/blob/main/non-mathlib/field_definition.lean#L23}{\color{blue}{\lstinline!square! function}}.

The code itself is available from Github.%
\footnote{\url{https://github.com/NicholasDyson/lean-polynomials}} 
It compiles with Lean 3.33 and, for the version that relies on \mathlib, a recent version of \mathlib.%
\footnote{Tested with \href{https://github.com/leanprover-community/mathlib/commit/b7593841620449def9435f0b9f3a1002afecff53}{Commit b7593841620449def9435f0b9f3a1002afecff53}.}
The files for the version based on \mathlib should be placed within the \lstinline!src/algebra! directory of \mathlib.
The code also works with the bundles%
\footnote{\url{https://leanprover-community.github.io/get_started.html\#maybe-a-couple-of-nights}}
distributed by the Lean Community.

\section{First proof version}
\label{sec:firstproof}

When first planning the project, we decided that we should aim to prove the relevant theorems from scratch, and without using any prewritten libraries.
This seemed likely to be more educational in terms of both learning to use \Lean effectively, and learning the underlying theory behind the formulae.

Our initial thinking was that, as the polynomial formulae operate on numbers, we should start by defining different types of numbers and proving theorems about them.
This was partly inspired by Kevin Buzzard's Natural Number Game~\cite{buzzard20}, which is the resource that we used to learn basic \Lean syntax and ideas.
This ``game'' starts with a definition of natural numbers (whereby a natural number is either zero or the successor of a natural number), and then walks the reader/player through proving some simple results based on that definition.
Based on this, our initial \Lean code also began with this same definition.
After defining various basic notions (addition, multiplication, associativity and commutativity of addition and multiplication,
distributivity of multiplication over addition, etc)
around natural numbers, a definition of integers was the first logical extension. 

\subsection{Implementations of the integers}
Our first implementation of integers was a type with two constructors---one that takes a natural number and represents that natural number,
and one that takes a natural number and represents the negation of that natural number.
\begin{lstlisting}
inductive myint : Type
| pos : mynat → myint
| neg : mynat → myint
\end{lstlisting}

While this may seem like the most obvious version to use (after all, an integer is either positive or negative),
it ended up being cumbersome to work with due to having two constructors.
This meant that whenever we wanted to work with a general integer, there needed to be a case analysis,
and as the number of possible cases was exponential in the number of integers involved in a given lemma, this quickly proved to be unworkable.

A better implementation,
which is also known more generally to mathematicians as the \emph{ring of differences}
(see, e.g.,~\cite[Proposition~4.2]{ELBASHIR2001277}),
turned out to be defining an integer as a pair of natural numbers,
where $(a, b)$ represents the integer $a - b$.
\begin{lstlisting}
inductive myint : Type
| int  : mynat → mynat → myint
\end{lstlisting}
It is not obvious why this is simpler, as now every integer has an infinite number of possible representations
(e.g., $-1$ can be represented as $(0, 1)$ or $(1, 2)$ or $(2, 3)$ etc),
but removing the need for multiple different constructors made working with the type much easier.
Various lemmas, in particular, the properties of a commutative ring, can then be demonstrated from these definitions.

\subsection{Implementing rational numbers}
Rational numbers are more interesting to implement, because by introducing a notion of division we open the possibility of invalid operations.  
Division by zero is not defined, but functions in a mathematical language like \Lean have to be total---they are not allowed to simply error and return nothing on certain inputs.  
There are various possible ways of handling the division-by-zero issue,
but the way we chose to implement it was to treat rational numbers with zero on the denominator as undefined values.
All functions would then be implemented so as to return undefined values if any of the inputs were undefined
(e.g. consider multiplication $\frac{a}{b} \* \frac{c}{d} = \frac{ac}{bd} $, and $bd = 0$ if and only if $b = 0$ or $d = 0$).
Thus, failure propagates through any sequence of operations---we cannot get a sensible value out of a formula that received nonsensical input.  
Accordingly, predicates (functions returning true or false) about rational numbers are defined to be True for undefined values.  

This has some somewhat undesirable consequences, such as sacrificing transitivity of equality
(if $a$ and $b$ are different defined values and c is an undefined value, $a = c$ and $c = b$ but $a \ne b$),
but importantly it allows lemmas to be proved without needing to specify that every individual argument is defined---if any of the inputs included division by zero,
then at least one of the values being tested for equality will be undefined, so the equality will be true.  

For example, consider the lemma for multiplication by zero
(\coqident{early_version}{rat_mul_zero_alt}):
\begin{lstlisting}
theorem rat_mul_zero_alt (a : myrat) :
  rat_eq rat_zero (rat_mul a rat_zero)
\end{lstlisting}
For defined $a$, $0a = 0$, and we can prove this via the normal methods.
If $a$ is undefined, then $0a$ is undefined, and so $0a = 0$ by the definition of equality as being true for undefined values.  
We can therefore prove that $0a = 0$ without needing to specify that $a$ is defined.
Equality is therefore defined as 
\[\frac{a}{b} = \frac{c}{d} \iff (ad = bc \,\lor\, b = 0 \,\lor\, d = 0).\]

\subsection{Square and cube roots: field extensions via dependent types}

As the quadratic formula---which we were aiming for as our first objective---requires square roots, we spent some time considering how to implement them.  
A true implementation of real numbers is awkward in most programming languages,
as by definition a real number is only fully specified by its infinite representation, but computers only have finite memory.  
However, as we had proven that the rational numbers are a field, we could implement each square root as a field extension.
Indeed, for a given rational $c$, numbers of the form $a + b\sqrt{c}$, where $a$ and $b$ are rational, are a field.  
It turns out that this is all we need\footnote{In fact, it is more than we need.
The quadratic formula never divides by a square root, so it is sufficient to show that $a + b\sqrt{c}$ is a commutative ring when the coefficients are a field.}
to implement the quadratic formula,
as we are only dealing with a single square root at a time,
namely that of the discriminant of the quadratic.
We implemented this by making use of dependent types.
\begin{lstlisting}
inductive rat_plus_sqrt (a : myrat) : Type
| rps : ∀ (a : myrat), 
          myrat → myrat → rat_plus_sqrt a
\end{lstlisting}

The above is similar to a constructor with three arguments, except that one of the arguments is moved into the definition of the type.  
Then, a value $(a, b)$ of type \lstinline!rat_plus_sqrt c! represents the number $a + b\sqrt{c}$.  
Because the $c$ argument is part of the definition of the type, however, we can only perform operations between two numbers with the same $c$ value.  
This means that operations like, for example, $a_1 + b_1\sqrt{c} + a_2 + b_2\sqrt{d}$ (where $c \ne d$) are not defined,
which means we can prove the field axioms for the set of values and operations generated by any given $c$.  

All-in-all, we were able to prove that (for quadratics with rational coefficients)
the quadratic formula always generates values that equal zero when substituted back into the quadratic equation (\coqident{early_version}{quadratic_formula_works}):
\begin{lstlisting}
theorem quadratic_formula_works (a b c : myrat) 
: iszero_rps 
     (discriminant a b c) 
     (quadratic_subst (discriminant a b c) 
                      a b c 
                      (quadratic_formula a b c))
\end{lstlisting}
This took a couple of months from start to finish, so it seemed like a different approach might be necessary if we were going to do the cubic and the quartic in a reasonable timeframe.  
Because having multiple possible representations of the same number (e.g. $\frac{1}{2} = \frac{2}{4}$) meant that we couldn't use the built-in identity type (denoted by $=$) in \Lean,
we had been having to define and use our own equality relation for integers and rational numbers.  
This meant that many of the built-in \Lean tactics did not directly apply, which massively slowed down writing code.  
If we have the goal of \lstinline!rat_eq (rat_add a b) (rat_add c b)! and the hypothesis \lstinline!rat_eq c a! (we used prefixes like \lstinline!int_!,
and \lstinline!rat_! to distinguish different implementations of the same operation, so \coqident{early_version}{rat_eq} is equality for rational numbers),
we would also need a hypothesis \lstinline!c_def : defined c! proving that the denominator of \lstinline!c! is not zero,
and then the required tactic with our lemmas would be 
\begin{lstlisting}
exact add_sub_rat c a b (rat_add c b) c_def h
\end{lstlisting}
This is already very wordy for such a simple change, and as the expressions got more complex it became vastly more so.
On some occasions we had even written Python scripts to write part of the \Lean code for us, as the rearranging was so awkward to do.
This did not seem like it would be sustainable to the more complicated world of the cubic and quartic.

\subsection{Comparing to the \mathlib library}
The main \Lean library \mathlib solves the aforementioned problems by using a different definition of rational numbers.  
The denominator is a natural number rather than an integer (this obviously has no effect on the representational power,
as the numerator can still be either sign), and two additional arguments are used.  
A rational number is then a numerator, a non-negative denominator, a proof that the denominator is not equal to zero,
and a proof that the numerator is coprime with the denominator.  
This removes the need to deal with undefined values, as division by zero is now impossible, and there is now only one way to represent each number.  
This means that, if the definition of integers allows for the use of the identity type, the representation of rational numbers will as well,
because of proof irrelevance.
However, at the time we were avoiding using or looking too much at \mathlib for previously-stated reasons.  
Defining a notion of ``coprime'' that was easy to work with and had enough supporting lemmas
to allow the operations on an equivalent implementation of rational numbers to be defined
would have probably meant a lengthy detour into areas of number theory with fairly tangential relevance to the overall project.  
Continuing to work explicitly with rational numbers seemed impractical, so we started again on a new version of the code.

\section{Working with general fields}
\label{sec:version_two}

\subsection{A definition of fields}
\label{sec:ver_two:field}

The first version of the proof had involved defining rational numbers and then proving that they were a field.
However, for the next version, we instead started with a definition of a general field, rather than starting with a specific implementation of a type and building towards the field axioms.
This meant using the \Lean notion of typeclasses, which work like interfaces in traditional programming.
For example, among other requirements we specify that,
for a type $F$ that implements the ``field'' typeclass, a binary operator of type $F \to F \to F$ that we call ``add''
and a proof that $\forall (x \: y : F)\, add \: x \: y = add \: y \: x$ must exist, which we call \lstinline!add_comm!.
The complete definition of a field that we use can be seen in the HTML documentation (\coqident{field_definition}{myfld}).

The most interesting line is 
\begin{lstlisting}
reciprocal : ∀ (x : a), (x ≠ zero) → a
\end{lstlisting}
This specifies that, for a field type \lstinline!a!, \lstinline!reciprocal! is a function that takes two arguments.
The first argument is a value \lstinline!x! of type \lstinline!a!, while the second argument is a proof that \lstinline!x ≠ 0!.
This means that we cannot generate the reciprocal of a value unless we are sure that it is not equal to zero,
which is of course in alignment with how division should work in mathematics.
This was easier to deal with than the way the previous version handled division by zero,
as now it was no longer necessary to handle two different possibilities (valid or invalid) for every reciprocal---the operation of division by zero was now impossible by definition and so did not need to be considered.
The \lstinline!mul_reciprocal! line then specifies that multiplying \lstinline!x! by \lstinline!reciprocal x _!, for any proof \lstinline!_! of \lstinline!x ≠ 0!, produces 1.

This definition of fields is enough to prove many facts without having to actually have a concrete implementation of any particular field,
with the additional advantage that the proofs have more generality.
In the previous version of the proof, we had proven that the quadratic formula is true in the fields of rational numbers (when extended with square roots).
This proof, however, had no innate ability to generalise to the fields of real numbers, complex numbers, etc,
as it relied on the implementation of rational numbers.
The new proof, by working with a generic field rather than any specific field, bypasses this problem.

\subsection{Defining square roots}
In the previous version, as we were explicitly working with rational numbers, roots had to be implemented as a field extension
(although without the reciprocal being defined, as there was no need to do so) because most roots are not rational.
A similar implementation of square roots in this version would look like:
\begin{lstlisting}
inductive item_plus_sqrt (f : Type) [myfld f] (rt : f) : Type
| ips : f → f → item_plus_sqrt
\end{lstlisting}
This code defines a type parameterised by an arbitrary field and a number within the square root.
We could use this to define operations of addition, negation, multiplication (and, if necessary, division)
of field elements and their square roots similarly to how we implemented them for rational numbers in the previous version.
We did do %
{a quick quadratic formula proof with this version}
as a proof-of-concept, but did not spend very long on this as this approach would be awkward to generalise to higher-degree polynomials.
The cubic formula~\eqref{eq:depr-cubic-sol} involves taking the cube root of a term containing a square root, so here it would be necessary to stack multiple field extensions on top of each other.
Notation would then quickly become awkward,
as it is necessary to specify the type of the base field whenever invoking a function that works on one of these extensions.
This would be undesirable, so we decided to use a different approach.

We instead require that the field itself should contain the roots.
This is simpler to implement, as it merely requires the definition of another typeclass of ``fields with square roots''
(\coqident{roots}{fld_with_sqrt}):
\begin{lstlisting}
@[class]
structure fld_with_sqrt (f : Type) [myfld f] : Type
sqrt : f → f
sqrt_mul_sqrt : ∀ (a : f), sqrt a .* sqrt a = a
\end{lstlisting}
These lines define a field with square roots as field with a function \lstinline!sqrt! such that for all \lstinline!a!, \lstinline!sqrt a * sqrt a = a!.

At a glance, this might seem too minimal for a complete definition of square roots,
as numbers besides zero have two square roots --- positive and negative.
However, we can obtain the other square root as the negative of the specified one:
(\coqident{roots}{negative_sqrt}),
\begin{lstlisting}
theorem negative_sqrt 
    (f : Type) 
    [myfld f] 
    [fld_with_sqrt f] 
    (a : f) 
 : .-  sqrt a .*  .-  sqrt a = a
\end{lstlisting}

We can also prove that no square roots other than these can exist.
One way to formulate this statement would be
\begin{lstlisting}[mathescape]
x * x = y 
⊢ x = sqrt y ∨ x = -(sqrt y)
\end{lstlisting}
However, while \Lean allows for classical logic we would prefer to stay constructive where possible (this second version of the overall proof,
since abandoning the original implementation of rational numbers, is entirely constructive), and this disjunction seems difficult to prove constructively:
as we did not specify anything about how the square root function should be implemented,
we cannot determine which of the two square roots $x$ is equal to.
However, the following version of the statement is constructively provable
(see \coqident{roots}{only_two_square_roots}):
\begin{lstlisting}
(x ≠ (sqrt y) ∧ (x ≠ -(sqrt y))
⊢ x * x ≠ y
\end{lstlisting}
%

\subsection{Quadratic formula}
\label{sec:ver_two:quadratic}

There is one more constraint that needs to be imposed on the definition of a field before we can define the quadratic formula.
The \textit{characteristic} of a field is the smallest number of times that $1$ can be added to itself to return to zero.
For example, in a field of characteristic $3$, $1 + 1 + 1 = 0$, or in a field of characteristic $5$, $1 + 1 + 1 + 1 + 1 = 0$.
If no such finite expression exists (so the field is infinite and no sum of $1$s returns to zero), we say the field is of characteristic $0$.
Recall that the quadratic formula~\eqref{quadr-sol} involves
dividing by 2, thus it is nonsensical if $2 = 1 + 1 = 0$,
\textit{i.e.}\ if the field has characteristic 2.
We implemented the additional constraint that the field in question is not of characteristic 2 with an additional typeclass (\coqident{numbers}{fld_not_char_two}):
\begin{lstlisting}
@[class]
structure fld_not_char_two (f : Type) [myfld f] : Type
not_char_two : myfld.one .+ myfld.one ≠ myfld.zero
\end{lstlisting}

Then, requiring that a type \lstinline!f! implements all three typeclasses (\lstinline!myfld!, \lstinline!field_with_sqrt!,
and \lstinline!fld_not_char_two!), is sufficient to define the quadratic formula
(\coqident{quadratic}{quadratic_formula}).
We implemented it for a general polynomial $ax^2+bx+c$
such that $a \neq 0$.

The proof of the validity of this formula is simple, and can take a couple of different forms.
Our initial proof (\coqident{quadratic}{quadratic_formula_works})
\begin{lstlisting}
theorem quadratic_formula_works 
    (f : Type) 
    [myfld f] 
    [fld_with_sqrt f] 
    [fld_not_char_two f] 
    (a b c : f) 
    (a_ne_zero : a ≠ myfld.zero) 
: quadratic_subst f (quadratic_formula f a b c a_ne_zero) a b c 
  = myfld.zero
\end{lstlisting}
was with the obvious approach of just substituting the value back into the original quadratic.
This is reasonably straightforward to prove via algebraic simplification:
\begin{align*}
a\left(\frac{-b + \sqrt{b^2 - 4ac}}{2a}\right)^2  + b\frac{-b + \sqrt{b^2 - 4ac}}{2a} + c &= \ldots = 0
\end{align*}
It is possible to translate the above approach into code in a general field, and it has the virtue of intuitive simplicity.
An essentially identical proof (that in the \Lean proof reuses a lot of the code to avoid unnecessary duplication) works for proving that the other solution
(where the square root is negated) is valid.
The weakness of this approach is that it does not allow for proving uniqueness, so another approach was needed.

The uniqueness proof (\coqident{quadratic}{quadratic_solution_unique})
\begin{lstlisting}
theorem quadratic_solution_unique 
   (f : Type) 
   [myfld f] 
   [fld_with_sqrt f] 
   [fld_not_char_two f] 
   (a b c x : f) 
   (a_ne_zero : a ≠ myfld.zero) 
 :  x ≠ quadratic_formula f a b c a_ne_zero 
  → x ≠ quadratic_formula_alt f a b c a_ne_zero 
  → quadratic_subst f x a b c ≠ myfld.zero
\end{lstlisting}
needs to take a similar form to the previous uniqueness lemma about how there can only be two square roots to a given number.
It was possible to prove this by reducing it back to the square root uniqueness lemma,
using a method similar to the ``completing the square'' method shown in \cref{sec:intro:quadratic}.

Finally, we also implement an alternative proof (\coqident{quadratic}{quadratic_factorize}):
\begin{lstlisting}
theorem quadratic_factorize 
    (f : Type) 
    [myfld f] 
    [fld_with_sqrt f] 
    [fld_not_char_two f] 
    (b c x : f) 
 : (x .+  .- (quadratic_formula f myfld.one b c myfld.zero_distinct_one)) 
   .* 
   (x .+  .- (quadratic_formula_alt f myfld.one b c myfld.zero_distinct_one)) 
  = 
   quadratic_subst f x myfld.one b c
\end{lstlisting}
that explicitly demonstrates that $(x - s_1)(x - s_2)$, for the two solutions $s_1, s_2$, is equivalent to the original quadratic.

\subsection{Cubic formula}
\label{sec:ver_two:cubic}

The quadratic formula, while not trivial, had been reasonably straightforward to implement
and prove once we were working with the general field rather than messing around specifically with our own implementation of rational numbers.
The derivation of the cubic formula is more complicated, as seen in \cref{sec:intro:cubic}.

To start with, a couple of new typeclasses are needed.
Obviously, a definition of cube roots is necessary to define the cubic formula, so the following typeclass (\coqident{roots}{fld_with_cube_root}) was used: 
\begin{lstlisting}
@[class]
structure fld_with_cube_root (f : Type) [myfld f] : Type
cubrt : f → f
cubrt_cubed : ∀ (a : f), (fld_with_cube_root.cubrt a) .* (fld_with_cube_root.cubrt a) .* (fld_with_cube_root.cubrt a) = a
\end{lstlisting}
This is a minimal definition of cube roots, and it is easy to see the similarities to the definition of square roots used earlier.
We can prove that the existence of the alternative cube roots follows from this definition and the definition of square roots.

We also need to define a field not of characteristic $3$ (as the formula involves dividing by multiples of three),
and the definition here is basically equivalent to the definition of a field not of characteristic $2$.

We can start the proof with a lemma (\coqident{cubic}{reduce_cubic_to_depressed})
\begin{lstlisting}
theorem reduce_cubic_to_depressed 
    (f : Type) 
    [myfld f] 
    [fld_not_char_three f] 
    (b c d x y : f) 
 : x = y .+  .- (b .* (third f)) 
  → 
    cubic_subst f myfld.one b c d x 
   = 
    depressed_cubic_subst f (square f b) .*  .- (third f) .+ c (twenty_seventh f) .* ((two f) .* (cubed f b) .+  .- ((nine f) .* (b .* c)) .+ (twenty_seven f) .* d) y
\end{lstlisting}
demonstrating the reduction of a cubic to a depressed cubic through a change of variable, as shown in \cref{sec:intro:cubic}.
However, the first attempt to implement the formula
that computes a solution to a depressed cubic did not work properly.
%

\subsubsection{First (failed) version}
\label{sec:ver_two:cubic:fst}

When solving the system of equations~\eqref{cubic_sims}
we initially thought that it would be easier to define the results as separate cube roots,
so that both $s$ and $t$ were defined in terms of the $c$ and $d$ values from the original equation as in~\eqref{eq:cubic-system-sol1} and \eqref{eq:cubic-system-sol2}.
This was how the results were presented in the first presentation of the proof we ever read,
so it seemed natural for the implementation at the time.
The reason why this approach failed is because of our minimal definition of cube roots. Indeed, we have no way of specifying what happens when we multiply two cube roots,
so we cannot prove any properties of $3st$. In particular, we cannot prove that $3st = c$.
Moreover, we cannot amend the definition of cube roots with a demand that the product of cube roots should equal the cube root of the product.
Indeed this assumption is in contradiction with the existence
of all square roots.
%
To see this, suppose we define a similar requirement on square roots, \textit{i.e.} we assume $\sqrt{a}\sqrt{b} = \sqrt{ab}$.
We then have
$-1 = \sqrt{-1}\sqrt{-1} = \sqrt{-1 * -1} = \sqrt{1} = 1$,
a contradiction (unless the field has characteristic 2).
We thus realised that we cannot define the product of square roots like this,
and a similar argument with one of the alternate cube roots of unity demonstrates that the same applies to cube roots.
This means that we have to express one of the two solutions
to the system of equations~\eqref{cubic_sims} as
$t = \frac{c}{3s}$ (or the converse where $s$ and $t$ are swapped),
as otherwise we have no way to prove that $3st = c$.

\subsubsection{Second version}
\label{sec:ver_two:cubic:snd}

The function \coqident{cubic}{cardano_formula}, which returns the solution to a depressed cubic given values for $c$ and $d$,
uses the same cube root~\eqref{eq:cubic-system-sol1} twice in the expression as in \cref{prop:depr-cubic-sol}.
As the proof no longer relies on any properties of the cube root function other than the one it is defined to have (that cubing it brings us back to the original number),
the function \coqident{cubic}{cardano_formula}
just takes the following additional arguments:
\begin{lstlisting}
cubrt_func : f → f
cubrt_func_nonzero : ∀ (x : f), x ≠ myfld.zero → cubrt_func x ≠ myfld.zero
cubrt_func_correct : ∀ (x : f), cubed f (cubrt_func x) = x
\end{lstlisting}
which are those used by the formula, rather than other functions to compute cube roots (like \lstinline!fld_with_cube_root.cubrt!).
This has the virtue that a proof of correctness of this new version of \lstinline!cardano_formula! can easily be generalised to all three of the cube roots,
simply by passing different functions as \lstinline!cubrt_func!.
Indeed, a function that multiplies the output of the function \lstinline!fld_with_cube_root.cubrt!
by one of the cube roots of unity will still satisfy the above requirements.

We did the proof of correctness (\coqident{cubic}{cardano_works})
\begin{lstlisting} 
theorem cardano_works 
    (f : Type) 
    [myfld f] 
    [fld_with_sqrt f] 
    [fld_with_cube_root f] 
    [fld_not_char_two f] 
    [fld_not_char_three f] 
    (c d : f) 
    (c_ne_zero : c ≠ myfld.zero) 
    (cubrt_func : f → f)
    (cubrt_func_nonzero : ∀ (x : f), 
       x ≠ myfld.zero → cubrt_func x ≠ myfld.zero)
    (cubrt_func_correct : ∀ (x : f), cubed f (cubrt_func x) = x) 
 : depressed_cubic_subst f c d 
       (cardano_formula f c d c_ne_zero 
          cubrt_func cubrt_func_nonzero 
          cubrt_func_correct)
  = myfld.zero
\end{lstlisting}
via substitution---simply demonstrating that if $x$ is equal to the value of the formula, then $x^3 + cx + d = 0$.

The proof of uniqueness first required defining the three different solutions.  
We did this by defining functions for the three different cube roots of a number,
and then creating new functions for the solutions that gave these cube root functions to the \lstinline!cardano_formula! function.
So, if we denote the value returned by \lstinline!fld_with_cube_root x! as $\sqrt[3]{x}$,
then the three cube root functions correspond to: 
\[
\sqrt[3]{x} 
\qquad
\frac{-1 + \sqrt{-3}}{2}\sqrt[3]{x}
\qquad
\frac{-1 - \sqrt{-3}}{2}\sqrt[3]{x} 
\enspace 
.
\]
These functions return different cube roots of $x$, and so passing them to the function \lstinline!cardano_formula!
will generate the three different solutions to a depressed cubic equation.
These are implemented as three separate functions: \coqident{cubic}{cardano_formula_a}, \coqident{cubic}{cardano_formula_b}
and \coqident{cubic}{cardano_formula_c}.
We then prove uniqueness of these solutions by showing that they factorize the original depressed cubic equation
(\coqident{cubic}{depressed_cubic_factorize}):
\begin{lstlisting} 
theorem depressed_cubic_factorize 
    (f : Type) 
    [myfld f] 
    [fld_with_sqrt f] 
    [fld_with_cube_root f] 
    [fld_not_char_two f] 
    [fld_not_char_three f] 
    (c d x : f) 
    (c_ne_zero : c ≠ myfld.zero) 
 : depressed_cubic_subst f c d x 
  = 
   factorized_cubic_expression f x (cardano_formula_a f c d c_ne_zero) (cardano_formula_b f c d c_ne_zero) (cardano_formula_c f c d c_ne_zero)
\end{lstlisting}
This entails uniqueness (of the form $x \ne s_1 \implies x \ne s_2 \implies x \ne s_3 \implies x^3 + cx + d \ne 0$) because the product of nonzero elements is nonzero.

As we had already shown how a cubic can be reduced to a depressed cubic,
the final results were easy to prove, with one example being
(\coqident{cubic}{cubic_formula_a_correct}):
\begin{lstlisting} 
theorem cubic_formula_a_correct 
    (f : Type) 
    [myfld f] 
    [fld_with_sqrt f] 
    [fld_with_cube_root f] 
    [fld_not_char_two f] 
    [fld_not_char_three f] 
    (a b c d : f) 
    (a_ne_zero : a ≠ myfld.zero) 
    (int_quantity_ne_zero : (three f) .* (a .* c) .+  .- (square f b) ≠ myfld.zero) 
 : cubic_subst f a b c d (cubic_formula_a f a b c d a_ne_zero int_quantity_ne_zero) 
  = 
   myfld.zero
\end{lstlisting}

\subsubsection{Final version}
\label{sec:ver_two:cubic:fin}
After implementing the above, it occurred to us that it might be possible to implement a more elegant version of the correctness proof
(if not the uniqueness proof) by using the simultaneous equations solution shown in \cref{sec:intro:cubic}---instead of the ``brute-force'' method of substituting the potential solutions back into the formula and simplifying it to zero.
As we were now only using a single cube root (that gets referred to twice) in the solution, it was possible to prove the solutions as shown earlier (see \coqident{cubic_alternate}{cardano_works}, with the same type as before modulo the definition of \coqident{cubic_alternate}{cardano_formula_new}).
While this derivation of the solution to a depressed cubic looked better, the rest of the proof
(reduction of a cubic to a depressed cubic, uniqueness, etc) is all structurally unchanged,
so we reused or left implicit a lot of code.



\subsection{Quartic formula}
\label{sec:ver_two:quartic}

The formalisation of the quartic formula actually ended up easier than the cubic formula, as we had the proof of the cubic solution to make use of.

First, observe that no new demands had to be placed on the fields in question: any field with square roots must also have fourth roots, and there is no such thing as a field of characteristic $4$,
as characteristics must always be prime numbers.

With these notions available, the proof can be implemented pretty similarly to how it is presented in \cref{sec:intro:quartic}.
First, we reduce the quartic to a depressed quartic (\coqident{quartic}{reduce_quartic_to_depressed}):
\begin{lstlisting} 
theorem reduce_quartic_to_depressed 
    (f : Type) 
    [myfld f] 
    [fld_not_char_two f] 
    (b c d e x u : f) 
 : x = u .+  .- (b .* (myfld.reciprocal (four f) _)) 
  → 
    quartic_expression f x myfld.one b c d e 
   = 
    depressed_quartic_subst f (depressed_quartic_squ_coeff f b c) (depressed_quartic_linear_coeff f b c d) (depressed_quartic_constant f b c d e) u
\end{lstlisting}
Afterwards, we reduce the depressed quartic to a product of quadratics (\coqident{quartic}{depressed_quartic_to_quadratic_product_solved}).
The full statement of this reduction is omitted here for reasons of space, but it has the form
\begin{gather*}
(d \ne 0 \land e \ne 0) \implies 
\\
 x^4 + cx^2 + dx + e = (x^2 + px + q)(x^2 + rx+ s)
\end{gather*}
where $p,q,r,s$ are as in \eqref{quartic-system}.

This approach to the proof has the advantage of entailing uniqueness
(\coqident{quartic}{depressed_quartic_solution_uniqueness})%
\footnote{Again, the full statement is omitted here, but it is of the same form as the previous uniqueness lemmas.}
in a simple way---once we have demonstrated that a quartic is equivalent to a product of two quadratics,
the proof of uniqueness of the two solutions to a quadratic entails the uniqueness of the four solutions to the quartic.

We had briefly investigated whether the original derivation of the solution by Lodovico Ferrari%
\footnote{Ferrari's solution is also described in Cardano's book~\cite{cardano}.}
might be easier,
but ended up choosing the quadratic product method because it looked simpler.
As it ended up being significantly more straightforward than the cubic proof, we think this was a correct decision.

The quartic formula is the one area where---in this version of the proof---we did slightly loosen our commitment to not using premade libraries/implementation,
by using \Lean's built-in \lstinline!nat! type.
Previously, all fixed constants in the formulae had been implemented from scratch,
using functions to return that given value for any general field type, such as (\coqident{numbers}{three})
\begin{lstlisting}
def three (f : Type) [myfld f] : f }
three f = myfld.one .+ (myfld.one .+ myfld.one)
\end{lstlisting}
This function defines \lstinline!three! as $1 + 1 + 1$, where 1 is the multiplicative identity in any given field.
Prior to the quartic, the largest constant that had to be involved in the formula was $27$, which is still fairly simple to define as $3 * 3 * 3$.
However, the quartic formula/algorithm has constants as high as 256, and implementing that as $2 * 2 * 2 * 2 * 2 * 2 * 2 * 2$ seemed awkward.
Therefore, we set up a function for powers of two that used the default \lstinline!nat! type
(\coqident{quartic}{pow_of_two}):
\begin{lstlisting}
def pow_of_two (f : Type) [myfld f] : N → f
pow_of_two f 0 = myfld.one
pow_of_two f n.succ = pow_of_two f n .* two f
\end{lstlisting}
This recursive function multiplies the field \lstinline!f!'s version of $2$ ($1 + 1$) as many times as the natural number given as input,
so \lstinline!pow_of_two f 8! computes to \lstinline!f!'s version of $256$.


\section{An alternate version using the \Lean standard library}
\label{sec:stdlib}

Implementing the previous version from scratch, starting from merely a definition of a field, was extremely educational to do.  
However, it seemed potentially worth it to do an alternate version (in the files 
\href{https://github.com/anonymousLeanDocsHosting/lean-polynomials/blob/main/mathlib/cubic_solution.lean}{\lstinline!cubic_solution.lean!} and
\href{https://github.com/anonymousLeanDocsHosting/lean-polynomials/blob/main/mathlib/quartic_solution.lean}{\lstinline!quartic_solution.lean!})
that made use of existing work, so as to be able to more fairly compare it to the Coq version.  
The \Lean standard library is called \mathlib~\cite{lean_mathlib}.  
The key benefit of using it for the new version is that we could make use of the various automation tactics that are included.  
These tactics were useful for doing the more mundane algebraic rearranging that we had previously had to do manually.  
The main factor slowing down the creation of this new version was looking through files for the proofs of elementary notions about fields.

We are planning to submit our proofs based on \mathlib for inclusion into the library, once the anonymity requirements allow for this.

\subsection {Discussion of stylistic differences in \mathlib} \label{sec:stdlib:compare}

There are two important differences in how notions about fields and roots are implemented between our version and \mathlib.
The first is that \mathlib defines $0^{-1} = 0$ (rather than making division by zero impossible as in our version),
and then requires a proof that $a \ne 0$ before the fact that $a^{-1} * a = 1$ can be used.  
This looks like nonsense to any mathematician, but the advocates of \Lean argue that it is functionally equivalent to, and simpler to work with than, a version like ours
where a proof of $a \ne 0$ is a prerequisite for taking the reciprocal of $a$~\cite{div_by_zero}.%
\footnote{This is not an uncontroversial decision, see the comments on \cite{div_by_zero}.}
The primary advantage is that it saves a bit of typing sometimes by not needing to reproduce proofs in every line that a reciprocal is used.  
The disadvantage of this approach is, of course,
that it moves us further away from the mathematical definition.

In addition, \mathlib has no definition of a field with square roots.  
Rather, the quadratic formula takes an extra argument \lstinline!s! for the square root, and the proof of correctness uses a proof
that $s$ is a square root of the discriminant.  
This may seem unnatural compared to just defining a notion of general square roots,
but it has the advantage of being able to prove the formula in a field where not every number has a square root.  
For example, consider the field of rational numbers.  
Most rational numbers do not have a rational square root, so it is impossible for this field to occupy the typeclass \lstinline!field_with_sqrt! that we defined in our version.
However, some rational numbers do have rational square roots, and if the discriminant of a quadratic is one of them
then the quadratic has a solution that can be arrived at without ever needing to use any irrational values.  
The \mathlib approach to defining the square root enables this to be represented.

Accounting for these differences, we were able to implement the proofs for \href{https://github.com/anonymousLeanDocsHosting/lean-polynomials/blob/main/mathlib/cubic_solution.lean}{the cubic} and \href{https://github.com/anonymousLeanDocsHosting/lean-polynomials/blob/main/mathlib/quartic_solution.lean}{quartic} formulae (and uniqueness of the solutions) in the \mathlib style.
The automation tactics such as \lstinline!field_simp! (which automatically rearranges equations in a field so as to move all terms over a common denominator) sped up parts of the work.
The actual logic of the proofs was generally similar to the earlier versions,
so between already understanding the logic of the proofs and being able to automate the more mundane rearranging,
the work proceeded very quickly in comparison to the previous version.

To appreciate the level of improvement in efficiency, consider that (measured by counting the number of commas in each file),
the non-mathlib \lstinline!cubic.lean! and \lstinline!quartic.lean! files have over 2000 tactics between them, while the \mathlib files are below 700.
Not every lemma from either version is exactly duplicated in the other
(and comparing the proofs for individual lemmas is of limited value as there is substantial variance in how useful the \mathlib additions were),
so these numbers are not perfectly comparable,
but it still suggests that use of \mathlib was able to at least halve the number of tactics required.

\section {Comparison to the Coq version}
\label{sec:coqcomp}

\subsection{Cubic}
\label{sec:coqcomp:cubic}

The first difference between the \Lean derivation of the cubic solution that we created and the existing Coq version~\cite{coq_proof} is that
the Coq version is specific to the field of complex numbers,
and it uses a specific function for $n$th roots based on De Moivre's formula,
instead of operating in a general field that contains the relevant square root and cube root and is not of characteristic two or three.
This makes the Coq proof slightly less general.
Examples of fields that do satisfy our conditions are
\begin{itemize}
\item the algebraic closure of a Galois field, and
\item the field of Puiseux series over some field.
\end{itemize}
On the other hand, it could be argued that a weakness of the typeclass approach we used in \Lean prior to \mathlib is that
we do not actually establish that any such structures exist.
For the quadratic formula, say, we assume the existence of a type that fits the definition of a field,
contains square roots, and is not of characteristic two.
The field of complex numbers obviously fits this description,
but as we did not implement that field, we have no formal proof that this is possible in the non-mathlib version.
The \mathlib library has field instances for the real numbers and the complex numbers, so this problem does not apply to that version of the proof.

A difference that is intrinsic to the proof assistants is that the automation tactics for simplifying expressions in a general field are implemented differently.
They both have a tactic \lstinline!ring! that attempts to automatically prove identities in rings and, of course, identities in fields that do not rely on division.
If division is involved, alternative tactics \lstinline!field! (in Coq) 
and \lstinline!field_simp! (in \Lean) 
\cite{lean_field}
that will attempt to simplify expressions with division in them can be used.
These field tactics both start by reducing the expression to the form $\frac{a}{b}$ ---
\textit{i.e.} putting everything over a common denominator if possible.
There are, however, two differences.
The tactic \lstinline!field! in Coq will then automatically try to prove the result using the functionality of the \lstinline{ring} tactic, while \lstinline{field_simp} does not.
This seems like a strange choice for Coq.
While one is often using \lstinline{field_simp} in preparation for closing the goal with \lstinline{ring}, this is not necessarily the case,
and there are points in the \mathlib proof where we chose to use \lstinline{field_simp} without immediately following it with \lstinline{ring} ---
cases where the equations were easier to work with with no/fewer division signs, but where the two sides were not yet trivially equal,
and so more work was needed before closing the goal with \lstinline{ring}.
We would not have been able to do this in Coq.

It is also worth noting how the Coq cubic proof does uniqueness and correctness in one go, simply by showing that
if we call the three solutions $s_1$, $s_2$, and $s_3$,
then $(x - s_1)(x - s_2)(x - s_3)$ is equal to the original depressed cubic expression.
This is how we approached the uniqueness proof in our version, and it is true that it is also valid for a correctness proof.
This is arguably a weakness of our implementation, as some of the code---the derivation and proof of correctness of the solution---is redundant from a certain perspective, as the final result could have followed from the correctness proof.
However, we think it was instructive to write the alternative derivation of the solutions.
Especially with the additional work described in \cref{sec:ver_two:cubic:fin},
we show the derivation of the formula rather than simply creating the formula ``from thin air'' and then proving/asserting that it is correct.

Both proof assistants have a command that displays all the axioms assumed by the proof of a lemma (and proofs referred to by that lemma, etc)
to demonstrate all the basic facts that a proof depends upon.
Using these demonstrates that the Coq proof assumes classical logic (the excluded middle)
\begin{lstlisting}
Classical_Prop.classic : ∀ P : Prop, P ∨ ¬P
\end{lstlisting}
while the \Lean versions (both the \mathlib version and the non-mathlib version) do not.
There does not seem to be any obvious way of checking where in the dependencies this requirement comes from ---
the Coq proof depends on many different libraries --- but we suspect it may come from the term \lstinline{Ceq_dec} that is referred to in the proof.
This seemingly determines whether or not two complex numbers are equal to each other.
Recall that
in the \Lean proof we required two separate formulae for the two cases where a given value was equal to zero or not.
However, the Coq version uses \lstinline{Ceq_dec} and the conditional (ternary) operator to combine these into one formula.

\subsection{Quartic}
\label{sec:coqcomp:quartic}

Allowing for these stylistic differences as described above, the Coq proof of correctness of the quartic formula uses a somewhat similar approach to the one we took,
using a function \lstinline{binom_solution} for the quadratic formula (where $a = 1$),
and demonstrating that the four roots of a quartic correspond to the roots of a pair of quadratics.
There is a minor difference in the implementation of \lstinline{binom_solution} compared to our own (non-\mathlib) quadratic solution which is interesting to note.
The definition of \lstinline{binom_solution} takes an extra argument \lstinline{n}, and multiplies the square root by $(-1)^n$.
This means that, if \lstinline{n} is even, the square root will be positive, whereas if \lstinline{n} is odd then the square root will be negative.
Therefore, the two roots of a quadratic can both be calculated from the same formula.

Besides this, the most substantial difference between the quartic proofs is that, similar to the cubic,
the Coq version combines everything relating to the formula---the reduction to a depressed quartic, the solution to the simultaneous equations, etc.---into a single large lemma, rather than using multiple lemmas to do each step.
This is partly due to a difference in the tools.
With the Coq tools we were using, it is possible to write a long proof,
and when making an edit then the compiler will only recompile sections below the point that has been edited.
On the other hand, the \Lean tools we were using seemed to start from the beginning of the proof of an individual lemma whenever any edit was made to that proof.
This makes it difficult to do very large single proofs,
as once a proof gets too long there can be a significant delay between adding a new line and seeing the new proof state as it stands after this line.%
\footnote{This delay sometimes seemed to be longer than the time taken to check the entire proof after it was finished. We do not know why this should be.}
This is a weakness of \Lean that caused us to occasionally split a proof into multiple parts even when there was no logical reason to do so,
such as with the lemma \coqident{cubic}{cardano_product_helper}---%
the comments on that lemma make it clear that it is simply the second half of the \lstinline{cardano_products} lemma (used for proving cubic uniqueness),
but was moved out to reduce the lag when editing.
This could feasibly be a Visual Studio Code problem rather than a \Lean problem,
and we have no information on whether it is reproducible by other people.

\section{Conclusion}
\label{sec:concl}

We have presented an implementation of the solutions of cubic and quartic polynomial equations, computer-checked in \Lean.
Compared to previous work on this topic, we use type classes to parametrize our solution by an arbitrary field equipped with necessary square and cube root functions.

We have not yet achieved the final goal we set out to achieve: a computer-checked proof that there is no closed formula for the roots of a general quintic polynomial.
However, we hope that the present work provides suitable foundations for formalizing this result.

\bibliographystyle{plain}
\bibliography{bibtex}


\end{document}